\documentclass[sn-mathphys,Numbered]{sn-jnl}% Math and Physical Sciences Reference Style

\usepackage{graphicx}%
\usepackage{multirow}%
\usepackage{amsmath,amssymb,amsfonts}%
\usepackage{amsthm}%
\usepackage{mathrsfs}%
\usepackage[title]{appendix}%
\usepackage{xcolor}%
\usepackage{textcomp}%
\usepackage{manyfoot}%
\usepackage{booktabs}%
\usepackage{algorithm}%
\usepackage{algorithmicx}%
\usepackage{algpseudocode}%
\usepackage{listings}%
\usepackage{caption}
\usepackage{subcaption}
\usepackage{threeparttable}
\usepackage{glossaries}
\usepackage{comment}

\newacronym{ssa}{SSA}{shared situation awareness}
\newacronym{smm}{SMM}{shared mental model}
\newacronym{ai}{AI}{artificial intelligence}
\newacronym{loa}{LOA}{level of abstraction}
\newacronym{EDL}{EDL}{entry, descent, and landing}
\newacronym{xai}{XAI}{explainable AI}
\newacronym{aa}{AA}{Automated/Autonomous Agent}
\newacronym{amp}{AMP}{Automated/Autonomous Mission Planner}
\newacronym{ndm}{NDM}{naturalistic decision making}

\begin{document}

\title[Article Title]{Improving Operator Situation Awareness when Working with AI Recommender Systems}

\author*[1]{\fnm{Divya} \sur{Srivastava}}\email{divya.srivastava@gatech.edu}

\author[2]{\fnm{J. Mason} \sur{Lilly}}\email{jmason.lilly@gmail.com}

\author[3]{\fnm{Karen M.} \sur{Feigh}}\email{karen.feigh@gatech.edu}

\affil*[1]{\orgdiv{Mechanical Engineering}, \orgname{Georgia Institute of Technology}, \orgaddress{\street{North Ave}, \city{Atlanta}, \postcode{30308}, \state{Georgia}, \country{USA}}}

\affil[2]{\orgdiv{Computer Science}, \orgname{Georgia Institute of Technology}, \orgaddress{\street{North Ave}, \city{Atlanta}, \postcode{30308}, \state{Georgia}, \country{USA}}}

\affil[3]{\orgdiv{Aerospace Engineering}, \orgname{Georgia Institute of Technology}, \orgaddress{\street{North Ave}, \city{Atlanta}, \postcode{30308}, \state{Georgia}, \country{USA}}}

\abstract{AI recommender systems are sought for decision support by providing suggestions to operators responsible for making final decisions. 
However, these systems are typically considered \textit{black boxes}, and are often presented without any context or insight into the underlying algorithm. 
As a result, recommender systems can lead to miscalibrated user reliance and decreased situation awareness. 
Recent work has focused on improving the transparency of recommender systems in various ways such as improving the recommender's analysis and visualization of the figures of merit, providing explanations for the recommender's decision, as well as improving user training or calibrating user trust.
In this paper, we introduce an alternative transparency technique of structuring the order in which contextual information and the recommender's decision are shown to the human operator. 
This technique is designed to improve the operator's situation awareness and therefore the shared situation awareness between the operator and the recommender system.

This paper presents the results of a two-phase between-subjects study in which participants and a recommender system jointly make a high-stakes decision.
We varied the amount of contextual information the participant had, the assessment technique of the figures of merit, and the reliability of the recommender system. 
We found that providing contextual information upfront improves the team’s shared situation awareness by improving the human decision maker’s initial and final judgment, as well as their ability to discern the recommender’s error boundary. 
Additionally, this technique accurately calibrated the human operator’s trust in the recommender. 
This work proposes and validates a way to provide model-agnostic transparency into AI systems that can support the human decision maker and lead to improved team performance.}


\maketitle

\section{Introduction}\label{intro}

In human-AI teaming, humans are paired with artificially intelligent ``partners" with the goal of leveraging the complementary strengths of each to improve task performance.

However, important problems arise when partnering with so-called ``black box" systems - AI systems whose decision making processes are either not available to or not easily interpretable by humans \cite{lipton-2017} - namely, that it is difficult or impossible to recognize erroneous output, or to know how to fix such errors.
Black box systems are particularly common in recommender systems where the model or algorithms underlying the \gls{ai} is often hidden from the human teammate \cite{Baker1995BreastCP, parmar2021}. 
Utilizing a black box \gls{ai} can lead to satisfactory results if the decision problem is well-characterized and simple or well aligned to the decision environment \cite{martignon2002fast}, or if the human is an expert in the task.
However, black box approaches are more prone to failure in complex settings and those in which the decision environment is challenging.
When the result of the \gls{ai}'s decision is unsatisfactory -- or even catastrophic -- there are typically no mechanisms in place to help users understand why that decision was suggested \cite{NTSB_accidentreport2014}.

Recommender systems have great potential to support humans in high-stakes domains. 
Significant work has sought to improve the quality of recommender systems and ameliorate the challenges associated with the complex and opaque algorithms inside.  
Efforts have included superior training for human operators \cite{yang2019study}, improved visualizations for proposed solutions \cite{becker2001visualizing, erra2011interactive}, ranking candidate solutions for the user to compare between \cite{monroe2019}, and attempting to explain the underlying reasoning for the suggested solution \cite{Dazeley_2021}.  
Each of these approaches have addressed challenges in niche classes of recommender systems, however identifying best human-AI interaction practices for recommender systems broadly remains an open problem.

This paper investigates an alternative method inspired by a combination of  \gls{ndm} and the growing literature showing the importance of having a \gls{smm} even between humans and artificial agents.  
Here we propose an approach designed to enhance the \gls{smm} between humans and recommender systems by focusing on improving the team's \gls{ssa} which can be defined as “a shared understanding of that subset of information that is necessary for [every teammate's] goals” \cite{endsley2003}. 
We wish to ground the human in the decision context and allow their judgment process to help supplement the joint decision.  
We hypothesize that providing contextual information, specifically designed to support the human's judgment and improve \gls{ssa} between the teammates, will positively impact overall performance, decrease over reliance, and help to calibrate trust. 
It will also provide one more tool to help improve the design of recommender systems.

\section{Background}\label{bg}
\subsection{Shared Mental Models} Mental models \cite{johnson-laird-1983} are internal representations of how and why a phenomenon occurs.
People create mental models of complex systems whenever they interact with one to facilitate their use of it \cite{norman-1983}.
 An extension of mental model theory is the concept of \textit{shared} mental models \cite{converse-1993}, i.e. mental models that are held in common across multiple individuals.
The key finding from this area regarding human-human teams is that if team members have similar mental models of their shared task and of each other, then they are able to accurately predict their teammates’ needs and behaviors.
\glspl{smm} enable high-performing teams in which everyone understands and anticipates the work of others in the team.
Teammates make decisions based on a common understanding of the state of the world which positively affects team performance.

Related work has applied this concept analogously to human-AI teams, seeking to foster mutual task-and-teammate understanding between a human and an \gls{ai} \cite{andrews2022}.
Work to improve the human's mental model of the \gls{ai}  \cite{bansal2019beyond} and vice versa \cite{fan-and-yen-2011, walsh2022} is ongoing, however, few empirical studies have attempted to quantify the development and effects of \glspl{smm} on human-autonomy teams.

Hanna and Richards' study found a meaningful correlation between the effects of trust and commitment to \gls{smm} development in teams of humans and intelligent virtual agents (IVAs) \cite{hanna-and-richards-2018}.
Better shared mental models were found to positively correlate with human trust in their artificial teammate.
In addition, teammate trust was found to significantly correlate with task commitment, which is found to significantly correlate with improved team performance.
However, these measurements were made via subjective self-assessments of \gls{smm} quality, and little research has attempted to establish a quantitative link on this topic.
This gap in literature can be attributed to limitations such as defining and implementing the \gls{ai}'s mental model and the methods used to elicit and measure the human-AI \gls{smm}.

\subsection{Recommender Systems \& \gls{ndm}}

Decision Support Systems (DSS) are a common application of \gls{ai} in various industries and have been around for decades, long before \gls{ai} methods were used to drive their internal algorithms.
They are designed to assist operators in decision making tasks with a goal of helping people to make better decisions faster.  
They do so by either simplifying the decision space or generating potential solutions to reduce the burden on human decision-makers.
A core benefit of decision support systems is their ability to encode expert-level information to support non-expert users.
\gls{ai}-driven recommender systems are a subset of DSS in which an \gls{ai}-based algorithm is utilized to recommend a course of action.  
Specific examples of \gls{ai} algorithms include machine learning, reinforcement learning, and deep neural networks.  
The process of decision making has been modeled in several ways \cite{dewey1910, simon1976, Kersten94}, but for the purposes of this paper, we will utilize the language of the well-established OODA loop \cite{hendrick-2009} which models the following cognitive processes:
\begin{itemize}
    \item Observation: the collection of data through sensory perception
    \item Orientation: the analysis and synthesis of data to form one's current mental perspective
    \item Decision: the determination of a course of action based on one's current mental perspective
    \item Action: the physical playing-out of the decision
\end{itemize}

The field of \gls{ndm} focuses on understanding, modeling, and improving how people make decisions and perform cognitively complex functions (such as observing and orienting) in demanding, real-world situations.
Several works within \gls{ndm} have shown that context and spending sufficient time and energy on the process of orienting or judging relevant information is influential to the decision part of the decision making process \cite{harringron-ottenbacher-2009, ben-akiva-2012, vazquez-diz-2019}.

%essentially involves 5 steps: recognizing a problem, gathering (relevant) information, generating possible solutions, making the decision, and finally evaluating the decision, which is only possible if there’s feedback.

In AI-advised decision making tasks, the \gls{ai} is responsible for the Observing and Orienting parts of the loop, and can also be used in the Decision part: The \gls{ai} gathers relevant information (Observe), and uses that information to generate possible solutions (Orient) or a judgment  of the decision that needs to be made (Decide).
The human’s role is mainly that of a safety checkpoint; the final decision (Decide) and implementation (Act) is the operator’s responsibility.
However, the black box nature of many systems limits appropriate understanding of what information the \gls{ai} is observing, and how the \gls{ai} orients that information to produce its outputs.
Because the human is not present in the Observation and Orientation parts of the decision making loop, this prevents the development of a team \gls{ssa} because often the human intentionally has little to no situation awareness.

The absence of a judgment phase for humans in the decision making process also leads to a common issue with recommender systems, which is the human's propensity to overrely on the system’s suggestion. 
Overreliance has been observed in a variety of domains \cite{wagner2018}, with both embodied \cite{wagner2016, booth2017} and nonembodied automated systems \cite{lee2004, bussone2015}. 
This can be attributed to a variety of explanations, though Wagner et al conclude that ``people might assume [automation] has knowledge [they] do not possess, viewing [automation]’s actions, at times mistakenly, as a direct reflection of the intentions of the developer, when in fact the [automation] may be malfunctioning or users may be misinterpreting its abilities. Even when presented with evidence of a system’s bad behavior or failure” \cite{wagner2018}. 
Whatever the underlying cause, overreliance can be mitigated when the human and the \gls{ai} have an accurate understanding of each other's tasks, roles, and responsibilities, and can make accurate predictions of each other's decision making capabilities (that is, if they had an accurate \gls{smm}).

\subsection{Known Contributions to Mitigate Challenges}

The field of \gls{xai} aims to address challenges brought upon by the black box barrier.
Some approaches attempt to “open” the black box by providing algorithmic transparency - insight into how an \gls{ai} agent processes information.
Much work in this area focuses on providing explanations for singular decision points \cite{lakkaraju2017, guidotti2018, Dazeley_2021}.
The most common types of explanation are global explanations, which explain the behavior of the model as a whole, and local explanations, which offer a specific explanation for that specific set of decision parameters.
An example of a global explanation would be a full set of heuristics or rules used to classify inputs into outputs, while a local explanation might offer specific rules to map inputs to outputs.

While common, this approach has limitations.
Explanations can take many forms \cite{guidotti2018}, and there is no consensus on what qualifies as a good or effective explanation.
For some \gls{ai} algorithms, explanations can be straightforward, such as displaying a decision tree or providing the features that triggered the decision.
For other algorithms, such as deep neural networks, the sheer volume of parameters and nodes used by the algorithm makes extracting a straightforward human-like explanation from them very difficult.
Additionally, the explanation given to the human is only useful if it is relevant to their existing mental models, and only if it succeeds at increasing the human’s understanding.
The given explanation needs to be communicated at a level of abstraction that makes sense to the individual user. 

If successfully implemented, algorithmic transparency can provide post-hoc context for an output to the decision maker. 
However, even if this is the case, the decision making process is still truncated for the decision maker because they do not \textit{Observe} any information and have to work backwards to \textit{Orient} the explanation with the AI's suggestion.

Other work has attempted to assist the human in evaluating the AI's suggestion by providing more interaction with the suggestion, or visualizing the suggestion differently \cite{becker2001visualizing, erra2011interactive}.
In simple algorithms, these techniques can increase understanding of the \gls{ai} model, but for complex decision making tasks, this alone may not be enough to support the human's judgment.

All of these techniques make valid contributions to the push to make human-AI decision making teams better, but as stand-alone techniques, they all truncate the decision making process which necessarily creates gaps in the decision maker's knowledge, and therefore, judgment.

More recent work has attempted to implement model-agnostic methods to interpret sophisticated machine learning models. 
Methods such as Local Interpretable Model-agnostic Explanations (LIME) \cite{ribeiro2016} and Shapley Additive Explanations (SHAP) \cite{lundberg2017} have made significant advances in increasing interpretability of complex algorithms. 
However, these methods are not very robust-- small changes in input features can lead to dramatic changes in explanations \cite{alvarez2018}. 
Additionally, SHAP requires exponential computational time \cite{kjersti2019}. 
While useful for fine-tuning algorithms, these methods are not suited to real-time, high-stakes decision making tasks.

Given that the transparency techniques truncate the human's decision making process and existing model-agnostic methods are limited in real-time capabilities, there is a need to support the human's real time decision making process in a way that does not rely on understanding the specific algorithms used by \gls{ai} recommender systems.
To this end, we propose our technique of providing contextual information of the task environment to better support the human's Orientation process when making a high-stakes decision.

\section{Methodology}

We conducted a two-phase between-subjects experiment in which participants were asked to work with an automated recommender system to complete a decision making task.
We manipulated the amount of contextual information the participants were provided, and we measured the effects on final task performance, trust in the recommender system, user workload, and the shared situation awareness between participants and the recommender.
Participants were recruited through an online recruitment platform (Prolific, www.prolific.co).
Individuals who were under 18, located outside of the USA, not proficient in English, and/or did not have normal or corrected-to-normal vision were excluded from the studies.
The experiments were conducted online and collected data from 90 participants in each study.
Participants were assigned randomly to a treatment group, and the order of scenarios shown to participants was balanced.

\subsection{Task Domain} 
Participants played the role of the commander of a craft in Mars orbit, tasked with finalizing the \gls{EDL} trajectory of a probe to a landing site.
Participants were told that they were aided by an AI Mission Computer that made a parallel evaluation of the proposed trajectory and offered agreement or disagreement with the participant's decision.
After seeing the AI's recommendation, the participant had the final call on whether to execute or abort the mission.
We imposed no time constraints to the task.
For control and reproducibility, though the system was presented to participants as ``intelligent'', its responses in each scenario were predetermined and fixed.

\begin{figure*}[!h]
     \begin{subfigure}{0.99\textwidth}
        \centering
        \includegraphics[width=\textwidth]{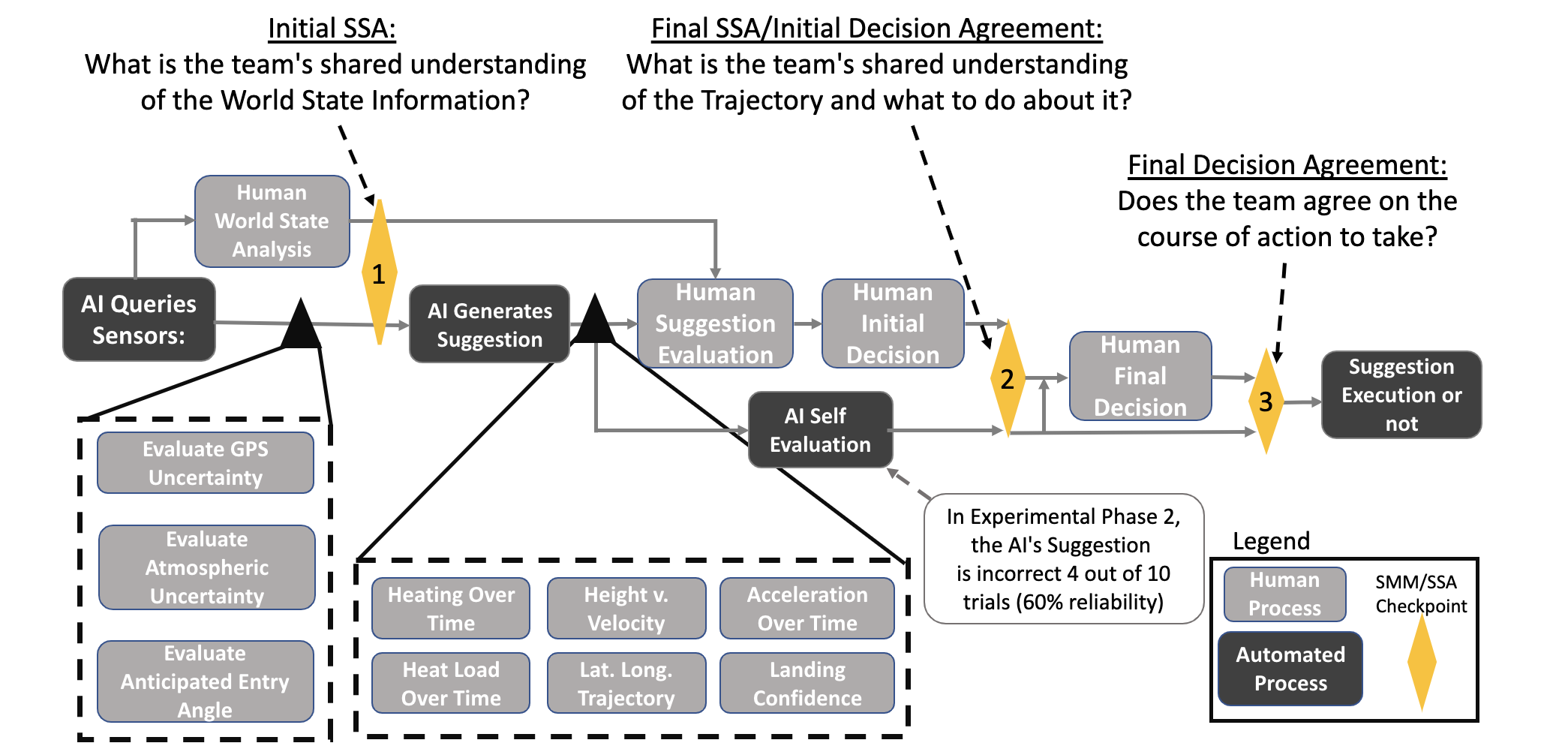}
        \label{fig:t1}
    \end{subfigure}

    \caption{Task Outline and Metrics for EDL Trajectory Planning}\label{fig:task}
\end{figure*}

The task is outlined in Fig.~\ref{fig:task}.
First, the two agents in the team (human participant and AI Mission Computer) independently assessed the state of the world, considering three factors: the orbital positions of GPS satellites, the locations of dust storms, and the atmospheric entry angle required by the craft's current orbital state.
Using this information, each evaluated whether the present world state permits a suitable landing (Initial Shared Situation Awareness Check represented by the yellow diamond \#1).

Then, the AI Mission Computer used the world state conditions to generate a possible trajectory for the probe's landing.
The team was shown a set of six figures of merit (FoM) that assessed the proposed trajectory: velocity vs.
altitude; heating rate, heat load, and acceleration vs time; latitude vs.
longitude; and landing confidence.
The FoM charts were shaded to indicate safe, risky, and dangerous thresholds, and the participants were instructed on how to interpret each.
Using the FoMs, the agents individually evaluated whether or not to execute the landing trajectory (Final Shared Situation Awareness Check represented by yellow diamond \#2).
The AI Mission Computer recommendation was made known to the participant following registration of their own.
The participant made a final decision on whether to execute or abort the mission in light of the AI's recommendation (Final Decision Agreement Point represented by yellow diamond \#3).

\subsection{Experiment Design and Task Procedure}

\begin{table}[h!]
    \small \centering 
    \caption{Design of Experiments}
    \begin{tabular}{@{} |c|c|c||c|c|c|}
      \hline
      Phase & IV & IV & DV & DV \\
     (Group) & World State   & Trajectory & World State & Trajectory  \\
        & Awareness & Awareness & Agreement  &  Agreement   \\
         \hline\hline
      1(1) &None & Obs       &   & X    \\
      1(2)& None & Int       &   & X    \\
      1(3)&Obs & Obs        &X  & X    \\
      1(4)&Obs & Int        &X  & X     \\
      1(5)& Int & Obs        &X  & X   \\
      1(6)& Int & Int        &X  & X    \\
      \hline
      2(1) &None & Obs       &   & X    \\
      2(2)& None & Int       &   & X    \\
      2(3)&Obs & Obs        &X  & X    \\
      2(4)&Obs & Int        &X  & X     \\
      2(5)& Int & Obs        &X  & X   \\
      2(6)& Int & Int        &X  & X    \\
      \hline
    \end{tabular}
    %\caption{Design of Experiments}
    \label{tb:TreatmentLevels2}
\end{table}

%This section details the experiment design and the tasks associated with each phase of the experiments.
Both experimental phases included two independent variables:  World State Awareness (3 levels) and Trajectory Awareness (2 levels). 
The second phase additionally manipulated the accuracy of the AI. In the first phase, the AI is correct in its evaluation of the trajectory (yellow diamond \#2 in Fig. \ref{fig:task}) in 100\% of the scenarios. 
In the second phase, this accuracy drops to 60\%, and the AI ``fails" in its suggested course of action when faced with a specific condition, which is unknown to the participant.
Each phase was designed as a 2x3 between-subjects design. We then compared between phases (making a 2x3x2 design, see Table \ref{tb:TreatmentLevels2}) to evaluate how the change in AI's reliability (100\% accuracy vs 60\% accuracy) affected the human's performance, team performance, workload, and trust. 
Participants were assigned to a single treatment in a single Phase, which was one of six combinations of the World State Awareness and Trajectory Awareness levels.  Participants completed 10 tasks within their treatment group. 

\subsubsection{World State Assessment}

\begin{figure*}[]
     \begin{subfigure}{0.32\textwidth}
        \centering
        \includegraphics[width=\textwidth]{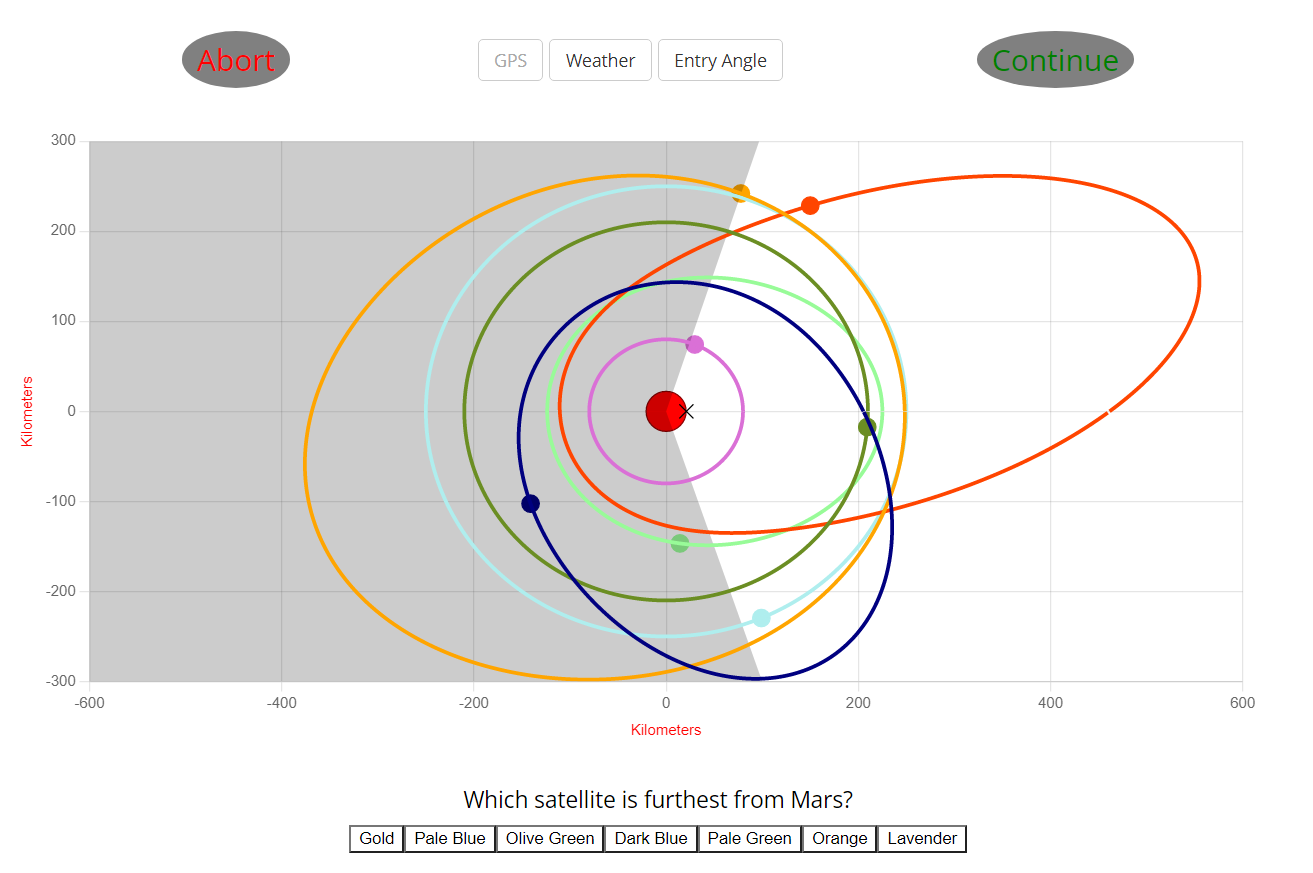}
        \label{fig:m1}
    \end{subfigure}
    \begin{subfigure}{0.32\textwidth}
        \centering
        \includegraphics[width=\textwidth]{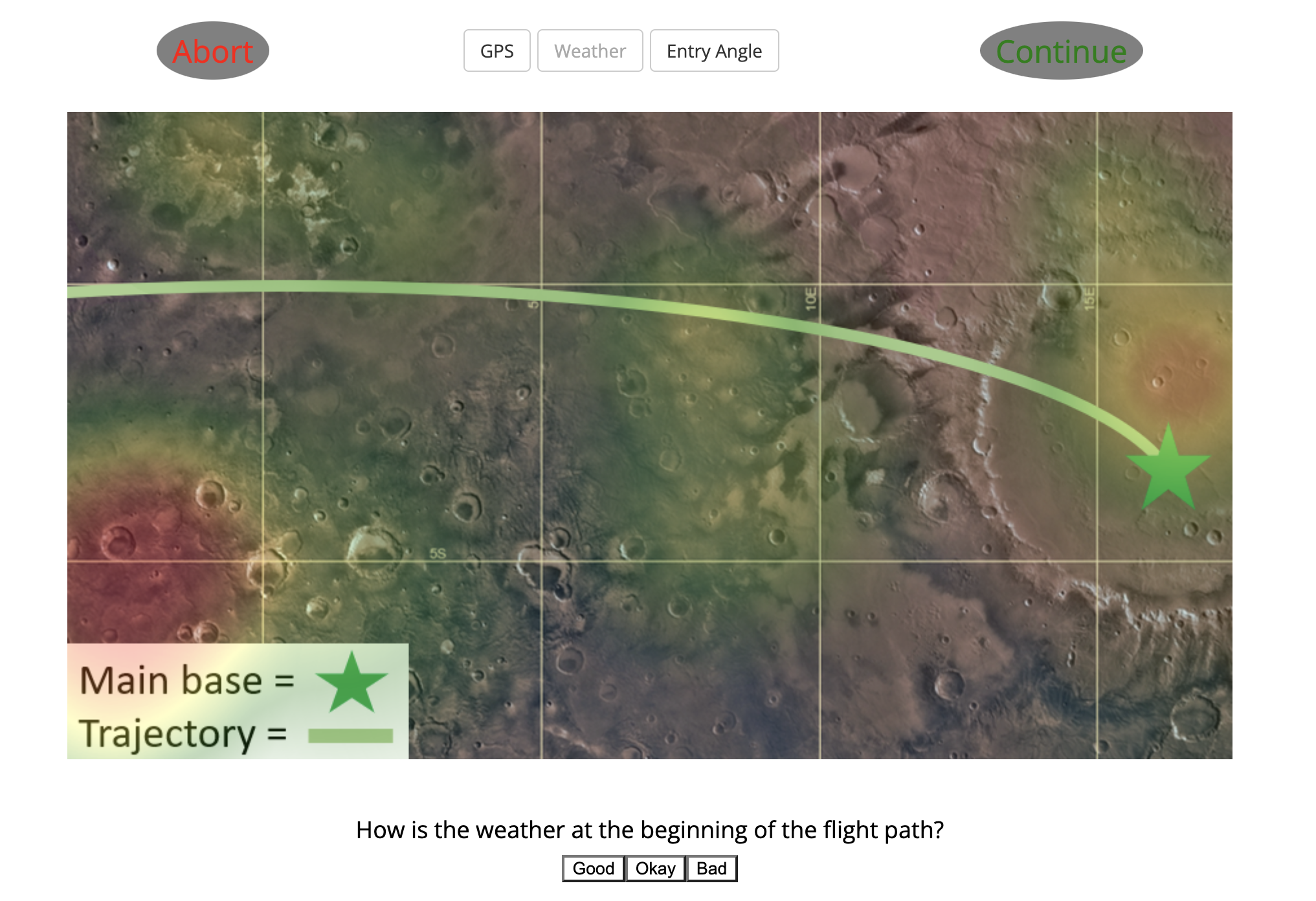}
        \label{fig:m2}
    \end{subfigure}
    \begin{subfigure}{0.32\textwidth}
        \centering
        \includegraphics[width=\textwidth]{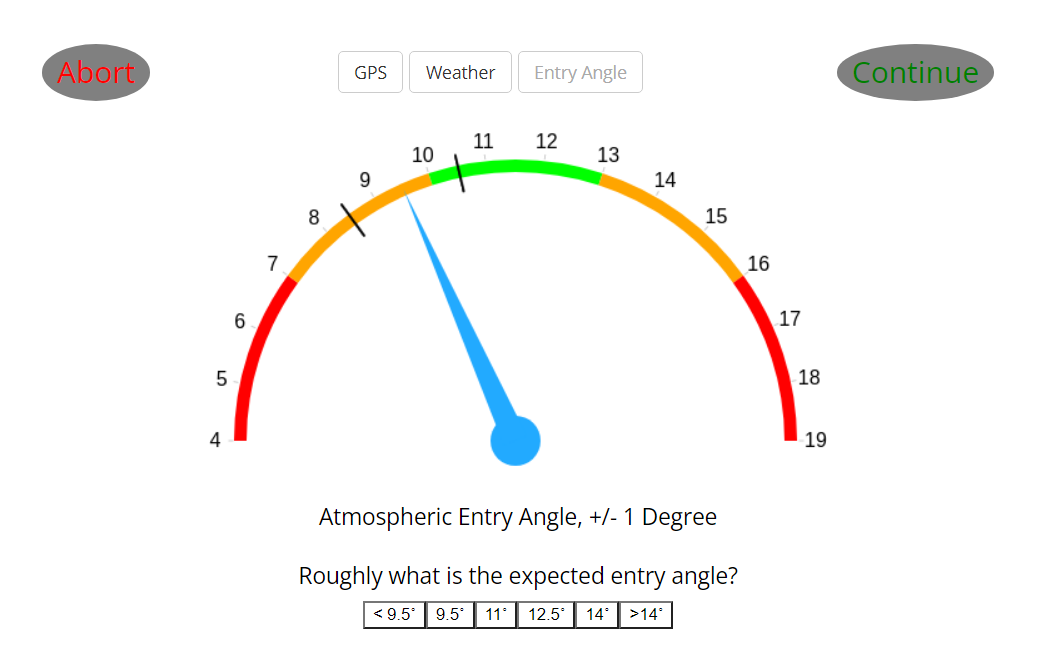}
        \label{fig:m3}
    \end{subfigure}

    \caption{World State Information from Left to Right: GPS, Atmosphere/Weather, Anticipated Entry Angle}\label{fig:ws_info}
\end{figure*}

In the first part of the task, the agents separately evaluated how risky the world state was (risky conditions vs safe conditions) in that scenario.
The World State Awareness independent variable had 3 levels: Observation-Only, Interactive, or Absent altogether.
The Absent mode was our control group, in which participants were not given any information about the world state (World State Awareness = None in Table \ref{tb:TreatmentLevels2}).
This lack of contextual information mimics current black box systems, and participants in this group started directly at the Trajectory Evaluation part of the experiment.
In the Observation-Only mode, participants viewed three information screens about the world state (Fig. \ref{fig:ws_info}).
They were not prompted for any further engagement with the information.
In the Interactive mode, participants viewed the world state information screens and were asked a multiple-choice question about each component, such as ``Which satellite is closest to Mars?" and ``How are the weather conditions near the landing zone?".
These questions were intended to increase the participant's situation awareness by forcing them to actively process each information source to a minimum degree.

After viewing and/or answering the question on each tab, participants were asked if the world state conditions were risky or safe enough to attempt a landing.
The AI Mission Computer prepared its own answer to the same question.
For implementation purposes, for all phases, these responses were always correct.  This was the first assessment of the participant and AI having a shared understanding, in this case of the world state.  
The participant was then informed of the AI's judgment of the world state conditions (and thus whether they are in agreement).
Then the experiment proceeded to the next step.

\subsubsection{Trajectory Evaluation}

 \begin{figure}[h]
    \centering
    \includegraphics[width=0.75\textwidth]{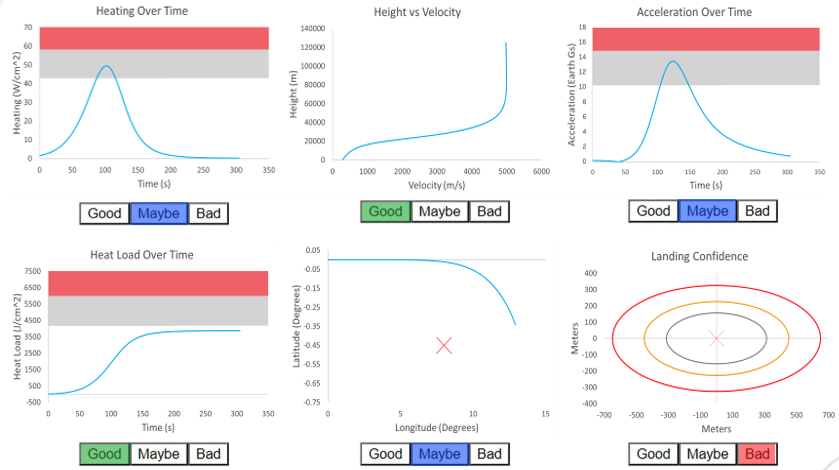}
    \caption{AI-generated trajectory with interactive questions}
    \label{fig:traj}
\end{figure}

Following the World State Awareness assessment (Treatment Groups 3-6 in both phases), a landing trajectory was presented, and the agents evaluated whether or not to execute it.
Participants were shown a spread of six figures of merit that characterized the proposed flight plan.
The Trajectory Awareness variable had two levels: Observation-Only or Interactive.
In the Observation-Only mode, participants reviewed the six figures of merit for the suggested trajectory and made an execute/abort decision.
In Interactive mode, participants were additionally asked to mark each of the six figures of merit as ``Good", ``Bad", or ``Maybe" according to that chart's specific risk factors (Fig. \ref{fig:traj}).
After observing (Groups 1, 3, 5) or interacting with (Groups 2, 4, 6)  each of the trajectory charts, participants were asked to decide whether to execute the trajectory.  This allowed us to get our second measure of shared understanding between the agents.  
\subsubsection{Final Assessment}
As in the World State Assessment, the AI Mission Computer shared its own evaluation after participants gave their assessment.   Then participants were asked to make a final decision.  This allowed us to understand how much the AI's own assessment influenced that of the participant.

\subsection{Experiment Procedure}
Participants were asked to complete a consent form and given a pre-experiment questionnaire to establish their baseline dispositional trust\cite{ithau}.
In both Phases, participants were assigned to one of six experimental treatments.
Instructional videos, tailored to the specific treatment, were used to train participants on the task and the AI Mission Computer.
No task-relevant experience was assumed.
%Participants were given descriptions of ideal conditions: For the world state, minimal dust weather, a middling entry angle, and at least four satellites; for the trajectory, charts below the ``risky'' threshold and small landing confidence ellipses.
Participants then completed 6 practice rounds of the mission planning task.
Those meeting a specific performance standard were allowed to proceed past the practice rounds.
Passing participants then proceeded to complete 10 trials of the mission planning task.
At the end of the experiment, participants completed two final questionnaires:  1) TLX workload \cite{hart2006} and 2) i-THAU trust assessment \cite{ithau}.

\subsection{Experiment Considerations and Limitations}
The ``AI'' behavior utilized in these studies is fully predetermined, with certain inputs (world state conditions) mapped to outputs (trajectory charts).
To simplify the analysis, the decision to execute or abort is decided beforehand, which means we can manipulate the accuracy of the AI's suggestion. 
In Phase 1, the AI's suggested course of action (whether to execute or abort a trajectory) is correct 100\% of the time. 
Designing the AI this way conflates team agreement with task performance, and as these two metrics are not realistically equivalent and 100\% correct performance from AI is unlikely, we manipulate the AI's reliability in Phase 2. 
In experiment 2, the AI is correct 60\% of the time, and fails in specific weather conditions. 
The decision to have the AI fail in a predictable way mimics real world situations in which algorithms that are trained on clearly labeled data and well-defined problems are successful when encountering similar situations, but fail when faced with noisy, degraded, or unfamiliar data. 
Additionally, by having the AI fail in a predictable manner, we were able to discern if providing transparency by way of contextual information helps the human understand the AI's limitations holistically (does the human pick up on the fact that the AI is giving inaccurate suggestions?) and specifically (under what circumstances does the AI fail?). 
These answers will inform the human's mental model of their AI partner, and thus their collaboration with it.

\subsection{Measures}
In each experimental phase, we first recorded the agreement between the participant's and the AI's initial judgment of the world state conditions (yellow diamond \#1 in Fig. \ref{fig:task}) and between their initial decisions of whether or not to execute the proposed trajectory (yellow diamond \#2 in Fig. \ref{fig:task}).
This served as a crude measure of the final shared situation awareness (\gls{ssa}) between the participant and the AI.
We computed this on a per participant basis across all 10 scenarios they experienced.
Second, we recorded the final agreement between the human and the AI after the AI's recommendation to execute/abort was revealed.
Third, we recorded the number of decisions per participant that changed from initial to final decisions and from there computed the percentage of initial disagreements that were resolved.
Fourth, for the interactive treatments, we were able to verify that the understanding of the human matched the understanding of the AI.

Additionally, responses were recorded from the three subjective questionnaires given to participants.
The pre-experiment questionnaire was an i-THAu trust assessment in which participants were asked to answer a series of statements about their Faith in Persons and Faith in Technology on a seven-point Likert scale from [-3:3].
A composite average of their answers informs their overall dispositional trust in these two categories.
The first post-experiment questionnaire was the NASA TLX workload assessment which measured overall cognitive workload.
The second was the i-THAu trust assessment in which participants responded to a series of statements about their experience of working with the automated system on a seven-point Likert scale from [-3:3].
For both i-THAu assessments, a rating of -3 indicated a lack of trust -- the subject didn't trust people/technology or depend on the AI to help them with the \gls{EDL} task, or they didn't understand the role of the AI.
Conversely, a rating of 3 indicated a high level of trust in people/technology, as well as high trust in and understanding of the AI.

\section{Results}
The results of both experimental phases obtained through objective and subjective measures are presented here. 
For all objective metrics, we performed a linear mixed effects analysis to determine the significance of the relationship between the metric and our independent variables.
The fixed effects were the World State Awareness, Trajectory Awareness, the average Dispositional Trust in People and the average Dispositional Trust in Technology.
We also included an interaction effect between World State Awareness and Trajectory Awareness, and intercepts for subjects as a random effect.
We performed an ANOVA for each fitted linear mixed-effects model.

\subsection{Experimental Phase 1: 100\% Accurate AI}
\subsubsection{Objective Metrics}
\begin{figure}[h]
    \centering
    \includegraphics[width=0.75\textwidth]{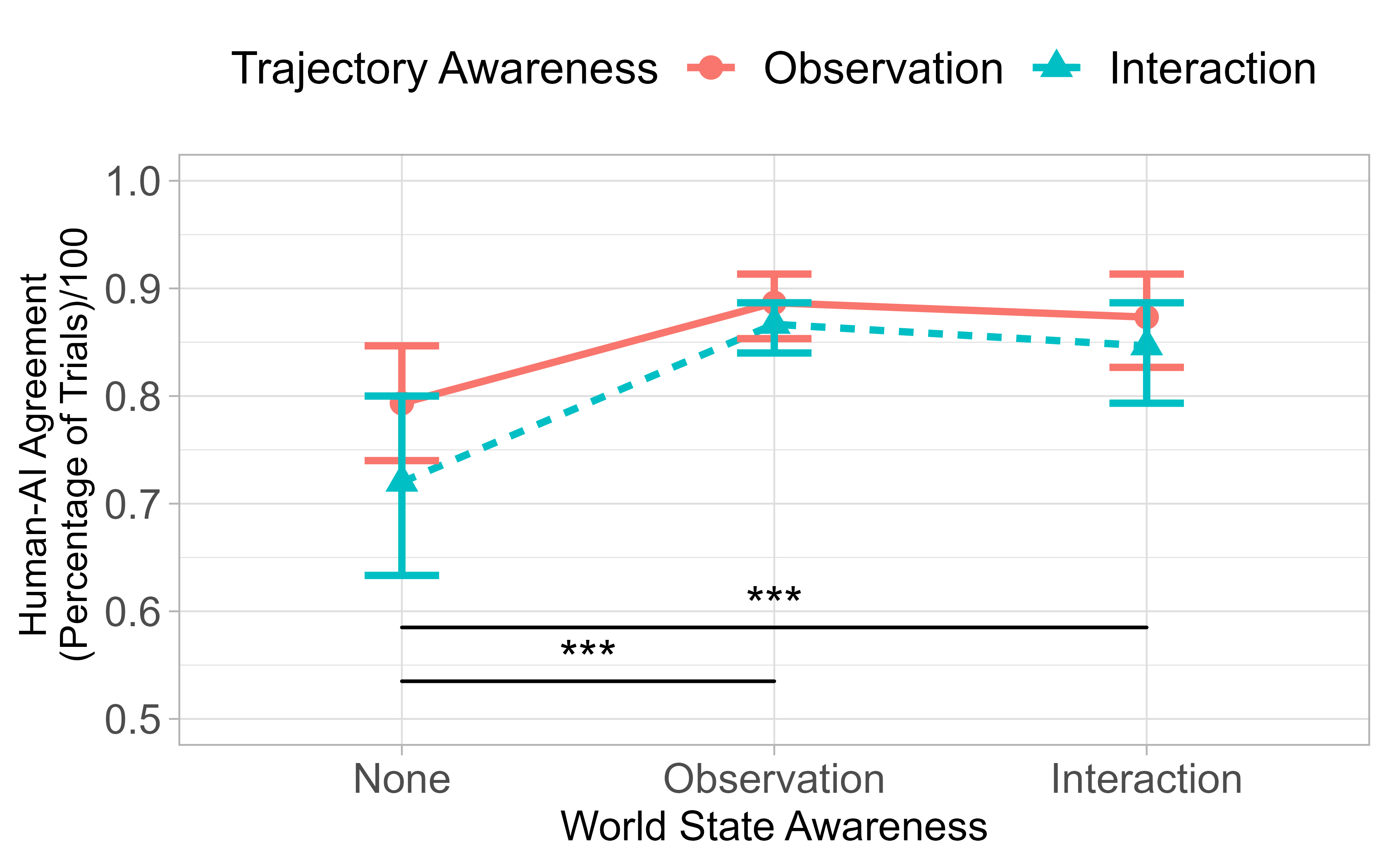}
    \caption{Final Shared Situation Awareness  [\%] Between Human and 100\% Accurate AI During Evaluation Phases}
    \label{fig:final_ssa}
\end{figure}

We first assessed the final shared situation awareness of the two teammates (yellow diamond \#2 in Fig.~\ref{fig:task}).  
Figure \ref{fig:final_ssa} shows the percentage of the initial agreement between the participant and the AI during the World State Assessment and Trajectory Evaluation phases.
Those who did not receive world state information were in agreement with the AI Mission Computer 60\% - 85\% of the time.
When world state information was available, the participant and AI Mission Computer tended to agree 80\% - 95\% of the time, and the variation reduced.
This indicates that being aware of the decision environment (the world state conditions) increased the shared situation awareness about the specific decision task, i.e. trajectory evaluation. And, as the AI was correct in this phase, the initial human performance also increased.  

\begin{table}[h]
\caption{\label{tab:SSA_anova}ANOVA for Final SSA}%
\begin{tabular}{@{}llll@{}}
\toprule
 & df, error & F & P \\
\midrule
WSAwareness & 2, 82 & 10.425 & 0.0001    \\
TrajAwareness & 1, 82 & 2.960 & 0.0891    \\
Faith in Tech & 1, 82 & 0.250 & 0.6181    \\
Faith in Persons & 1, 82 & 0.596 & 0.4424    \\
WS-A:Traj-A & 2, 82 & 0.650 & 0.5248    \\
\botrule
\end{tabular}
\end{table}

Table~\ref{tab:SSA_anova} presents the  ANOVA for the fitted linear mixed-effects model of the initial decision agreement between the participant and the AI Mission computer before the AI's decision was revealed (yellow diamond \#2 in Fig. \ref{fig:task}).
The results indicate that the World State Awareness significantly impacts  the initial agreement between the participant and AI.
No other effects were significant.

\begin{figure}[h]
    \centering
    \includegraphics[width=0.75\textwidth]{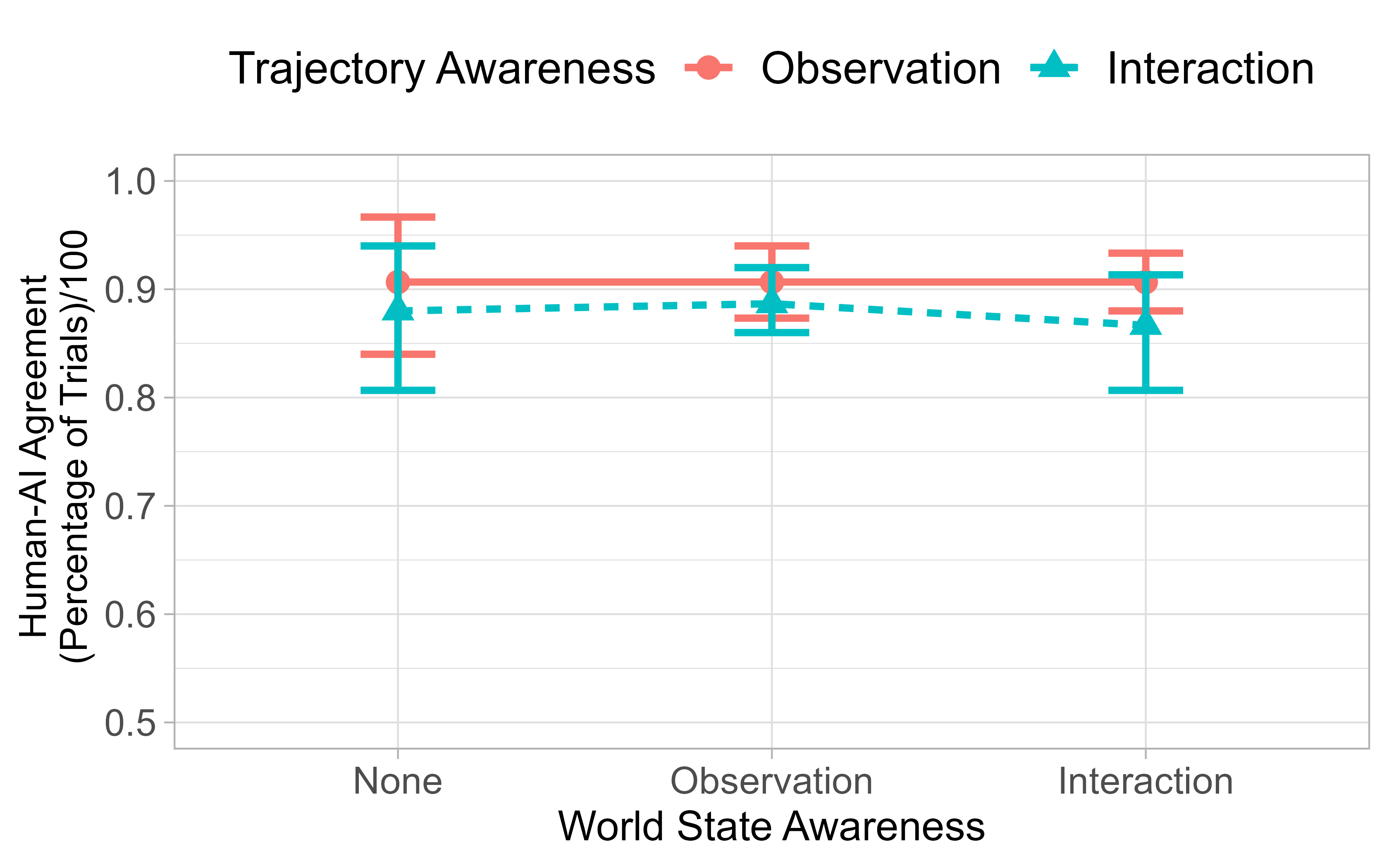}
    \caption{Final Agreement [\%] Between Human and 100\% Accurate AI}
    \label{fig:final_agr}
\end{figure}

\begin{figure}[hb!]
    \centering\includegraphics[width=0.75\textwidth]{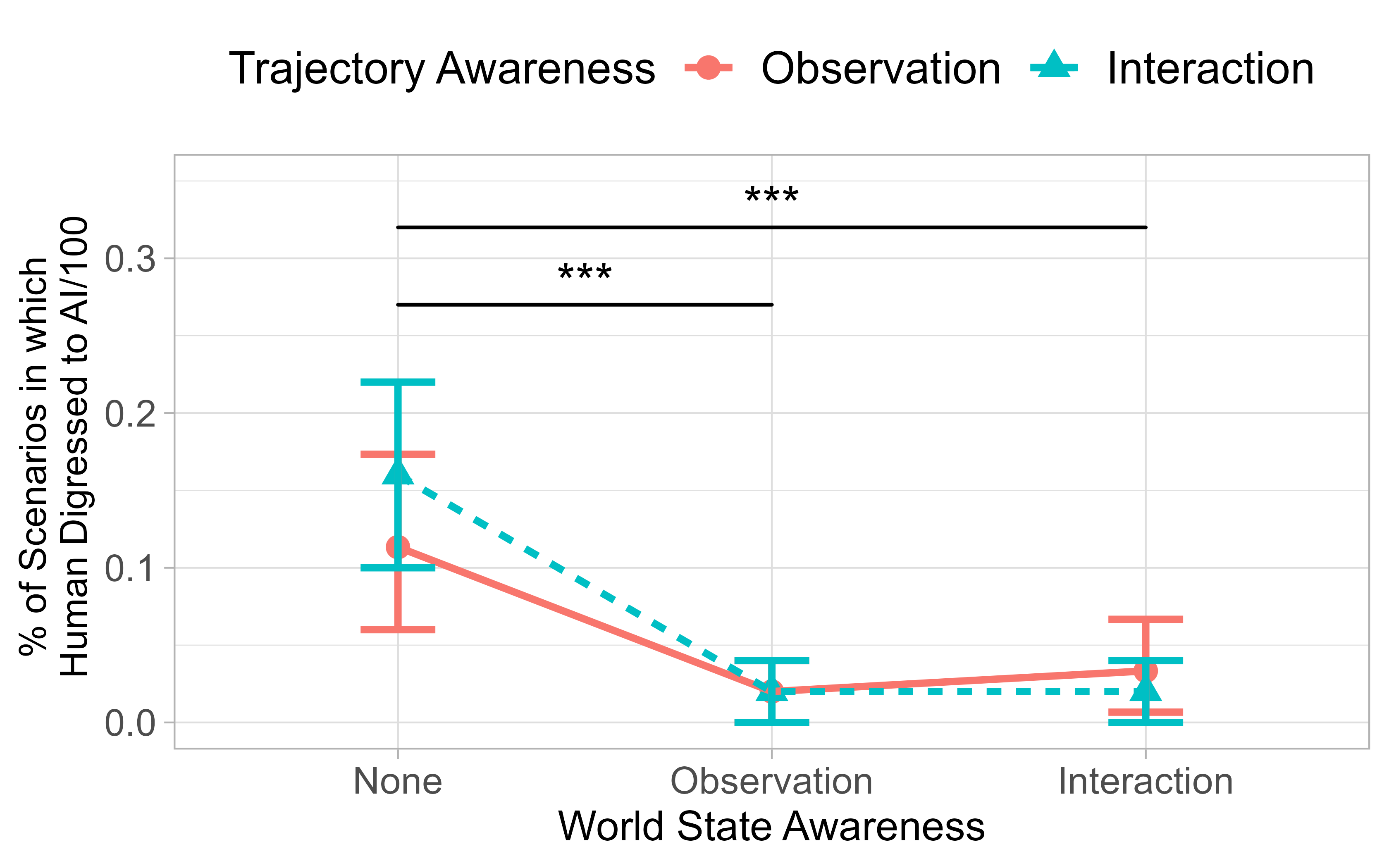}
    \caption{Resolved Team Disagreements [\%] Post-AI Suggestion}
    \label{fig:sway}
\end{figure}

We also analyzed how the overall team performed by assessing the team's final agreement (after the AI reveals its decision, shown as yellow diamond \#3 in Fig~\ref{fig:task}), seen in Figures \ref{fig:final_agr} and \ref{fig:sway}.
The average final agreement was around 90\% for all levels of World State Awareness but the variance decreased as World State Awareness increased.  Here it seemed an accurate AI helped the participants who did not have World State information perform as well as those who did.
A linear mixed effects analysis for this metric revealed that none of the fixed effects were statistically significant in predicting the final agreement between the team (Table \ref{tab:final_agr_anova}).

\begin{table}[h]
\caption{\label{tab:final_agr_anova} ANOVA for Final Agreement}%
\begin{tabular}{@{}llll@{}}
\toprule
 & df, error & F & P \\
\midrule
WSAwareness & 2, 82 & 0.076 & 0.9265    \\
TrajAwareness & 1, 82 & 1.844 & 0.1782    \\
Faith in Tech & 1, 82 & 0.357 & 0.5518    \\
Faith in Persons & 1, 82 & 0.483 & 0.4888    \\
WS-A:Traj-A & 2, 82 & 0.030 & 0.9706    \\
\botrule
\end{tabular}
\end{table}

\begin{table}[h]
\caption{\label{tab:sway_anova} ANOVA for Resolved Team Disagreements}%
\begin{tabular}{@{}llll@{}}
\toprule
 & df, error & F & P \\
\midrule
WSAwareness & 2, 82 & 18.95 & $<$0.0001    \\
TrajAwareness & 1, 82 & 0.408 & 0.5244    \\
Faith in Tech & 1, 82 & 0.004 & 0.9508    \\
Faith in Persons & 1, 82 & 0.933 & 0.8565    \\
WS-A:Traj-A & 2, 82 & 1.183 & 0.3115    \\
\botrule
\end{tabular}
\end{table}

A well known challenge in human-automation/AI teams is that  humans can often overrely on their computational teammates. Figure \ref{fig:sway} depicts the percentage of scenarios in which the participant reconsidered their initial decision to align with the AI after seeing what the AI suggested during the evaluation phases.
%We performed a linear mixed effects analysis of the relationship between this gap in agreement between the teammates and the aforementioned independent variables.
%We then conducted an ANOVA on this model (Table \ref{tab:sway_anova}).
The results from the ANOVA for this metric (Table \ref{tab:sway_anova}) indicate that the World State Awareness is statistically significant in predicting when the participant will change their decision to align with the AI's suggestion.  
No other effects were significant.
Based on Figures \ref{fig:final_agr} and \ref{fig:sway}, we can see that those who did not have access to the World State conditions had a tendency to be swayed by their AI partner to make the correct decision, while those who did observe or interact with the world state conditions had higher initial agreement with the AI.  
In this phase, agreement with the AI is equivalent to correct assessment, and so the tendency to agree with the AI may be seen as positive behavior.

\subsubsection{Subjective Metrics}

\begin{figure}[ht!]
    \centering
    \includegraphics[width=0.8\textwidth]{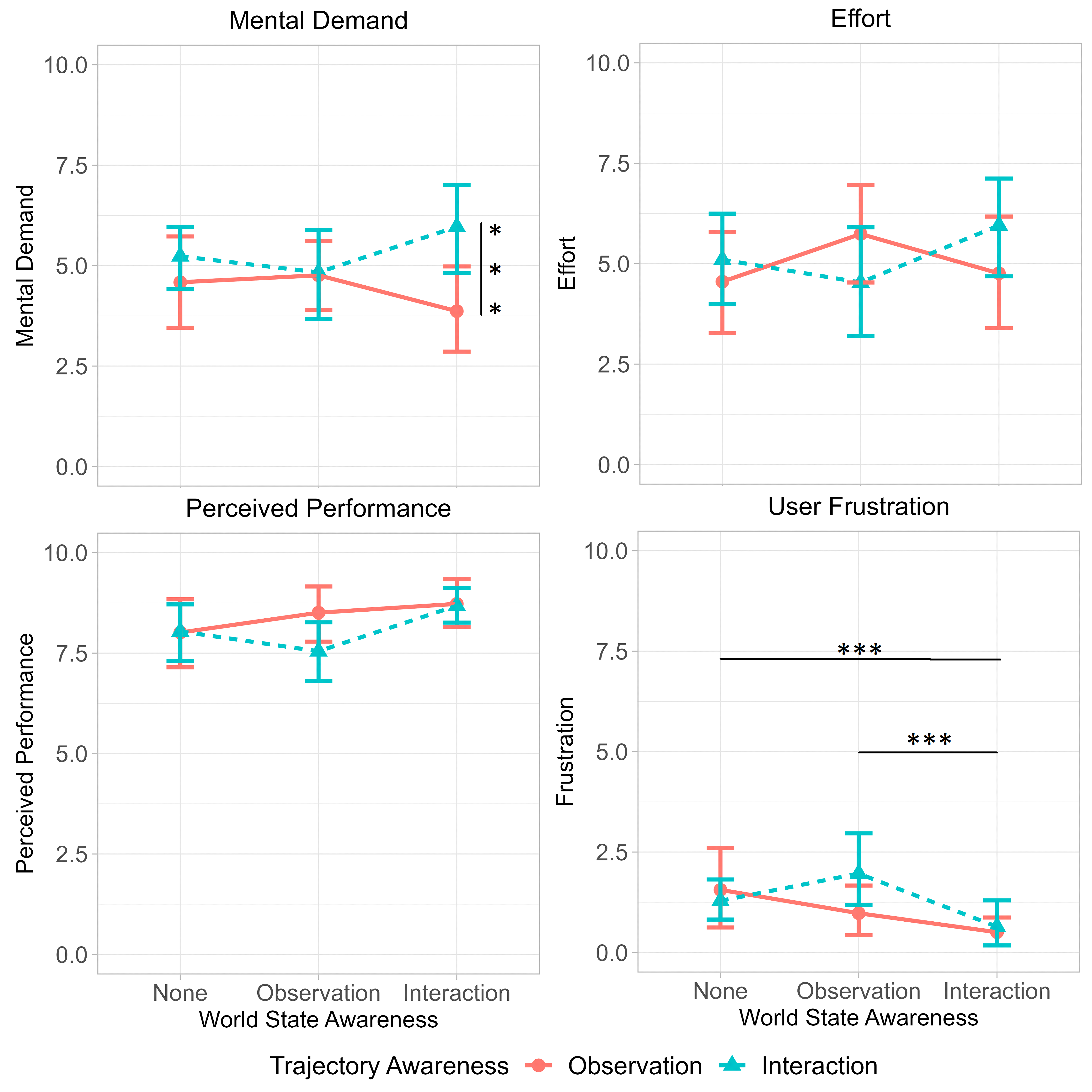}
    \caption{Composite Results of NASA TLX Questionnaire}
    \label{fig:exp1_tlx}
\end{figure}

To determine if adding the extra tasks of allowing or requiring participants to be aware of the World State and Trajectory Assessments impacted the user's mental workload, we conducted a subjective mental workload assessment (NASA TLX) at the end of the experiment.  
Figure \ref{fig:exp1_tlx} depicts the results of 4 of the 6 the NASA TLX subscales.
%For all metrics except perceived performance, higher values indicate a higher workload on the participant.
%For perceived performance, higher values indicate a better perception of their own performance.
Overall, increasing the participant's situation awareness required more effort of them but ultimately was less frustrating and resulted in increased perceived performance.
Results also show that those who interacted with the figures of merit (as opposed to merely observing them) during the Trajectory Evaluation phase tended to experience more mental demand and frustration, and perceived themselves as less successful in their performance.

\begin{figure}[h]
    \centering
    \includegraphics[width=0.75\textwidth]{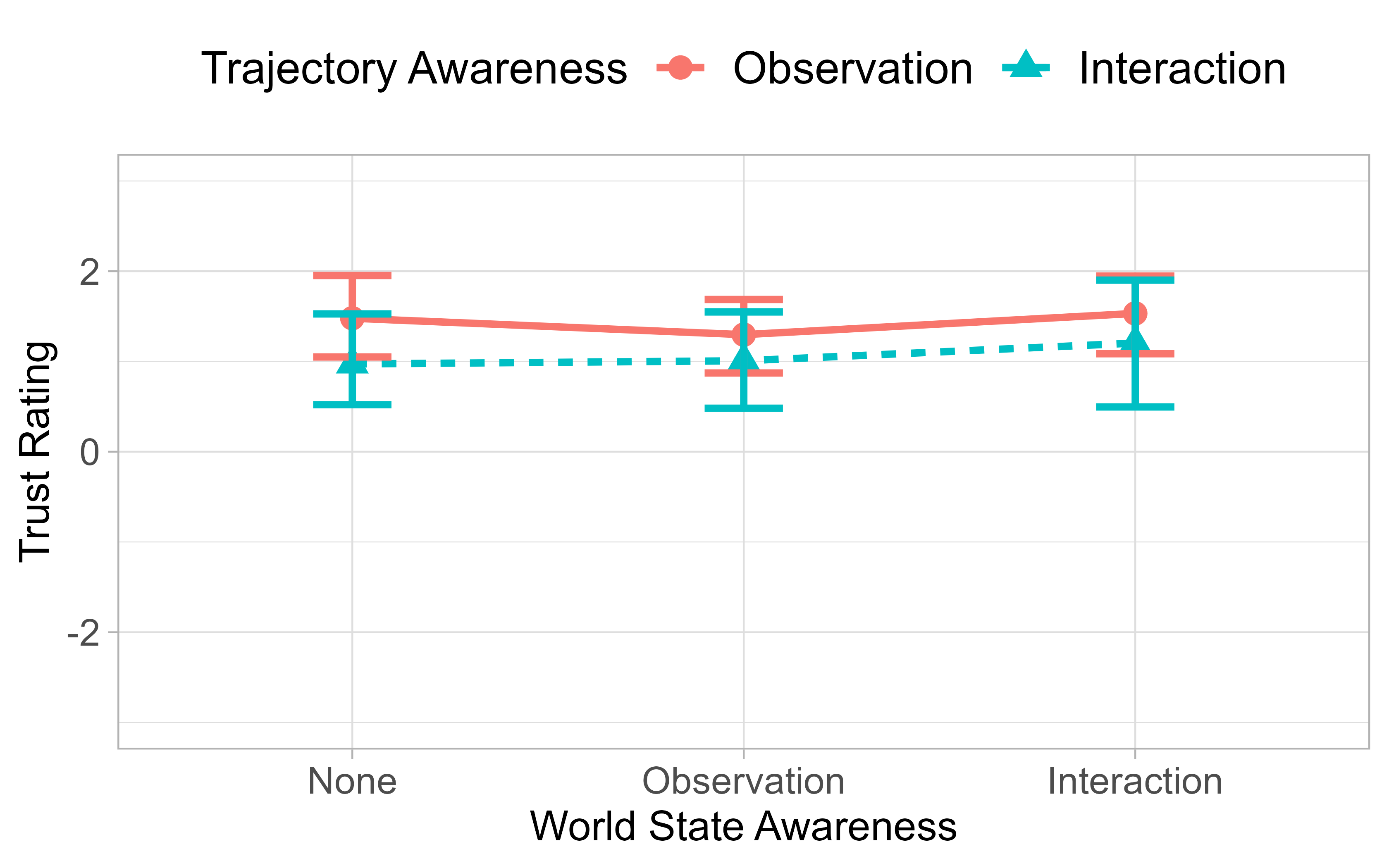}
    \caption{Post-Experiment i-THAu Metric of Human's Trust in AI's Capabilities}
    \label{fig:exp1_trust_capability}
\end{figure}

The i-THAu assessment measures different aspects of trust between humans and automation\cite{ithau}.  We used it for both pre- and post- trust assessments. The pre-experiment i-THAu survey measures dispositional trust, that is an individuals’ propensity to trust other people or technology. Our analysis revealed no impact of these trust factors on any objective measures such as SSA (Table \ref{tab:SSA_anova}), final team agreement (Table \ref{tab:final_agr_anova}), etc., nor were they found to influence subjective measures such as cognitive workload or learned trust. The post-experiment i-THAu assessment measures situational and learned trust. Figure  \ref{fig:exp1_trust_capability} shows the human's trust in the AI's capabilities to complete the task, which is measured following all 10 data collection scenarios. Positive values (+3)  indicate more trust in the AI's capabilities, while negative values (-3) indicate less trust in the AI's capabilities. While the presence of and interaction with world state information did not influence people's trust in the AI's capabilities, there was (non-statistically significant) reduced level of trust from those who interacted with the trajectory's figures of merit vs those who merely observed them. 

\subsection{Experimental Phase 2: 60\% Accurate AI}
In this phase of the experiment, the accuracy of the AI's assessment of the trajectory was reduced from 100\% to 60\%.  60\% was chosen as it is below the standard 70\% threshold at which human participants begin to be impacted by poorly performing automation \cite{wickens-and-dixon2007}.

\subsubsection{Objective Metrics}
\begin{figure}[h]
    \centering
    \includegraphics[width=0.75\textwidth]{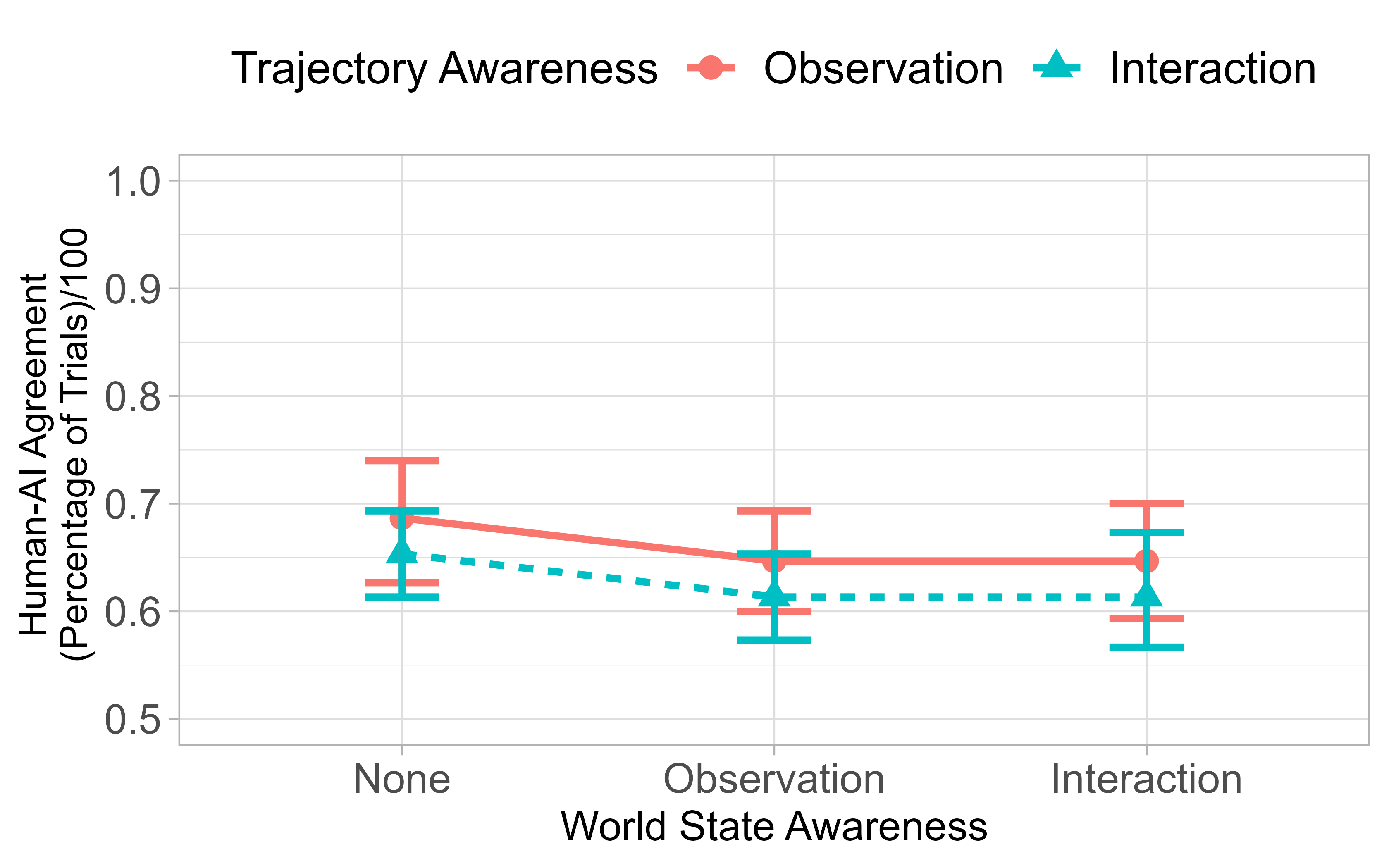}
    \caption{Final Shared Situation Awareness  [\%] Between Human and 60\% Accurate AI During Evaluation Phases}
    \label{fig:final_ssa_exp2}
\end{figure}

As before, we assessed the final shared situation awareness of the two teammates (yellow diamond \#2 in Fig~\ref{fig:task}).  
Ideally the human would only agree with their AI partner 60\% of the time. This would indicate that the human can recognize a limitation in their AI partner and adjust their mental model and actions accordingly to ensure successful team performance. 
Figure \ref{fig:final_ssa_exp2} shows the percentage of the final SSA between the participant and the AI during the World State Assessment and Trajectory Evaluation phases.
Those who did not receive world state information tended to be in agreement with the AI Mission Computer 60\% - 75\% of the time.
Once world state information was introduced, the participant and AI Mission Computer tended to agree 55\% - 70\% of the time. Additionally, those who interacted with the trajectory figures of merit trended closer to the ideal 60\% agreement than those who merely observed, and the variation was reduced.
This indicates that being aware of the decision environment (the world state conditions) helped to calibrate the accuracy of the team's shared situation awareness about the specific decision task, i.e. trajectory evaluation.

\begin{table}[h]
\caption{\label{tab:SSA_anova_exp2} ANOVA for Final SSA (Phase 2)}%
\begin{tabular}{@{}llll@{}}
\toprule
 & df, error & F & P \\
\midrule
WSAwareness & 2, 82 & 1.577 & 0.2127    \\
TrajAwareness & 1, 82 & 2.465 & 0.1203    \\
Faith in Tech & 1, 82 & 0.020 & 0.8871   \\
Faith in Persons & 1, 82 & 1.181 & 0.2804   \\
WS-A:Traj-A & 2, 82 & 0.004 & 0.9962    \\
\botrule
\end{tabular}
\end{table}

Table~\ref{tab:SSA_anova_exp2} presents the ANOVA for the fitted linear mixed-effects model for final SSA.
The results indicate that no factors were statistically significant in predicting the final SSA between the participant and AI when the AI was 60\% accurate.  

\begin{figure}[h]
    \centering
    \includegraphics[width=0.75\textwidth]{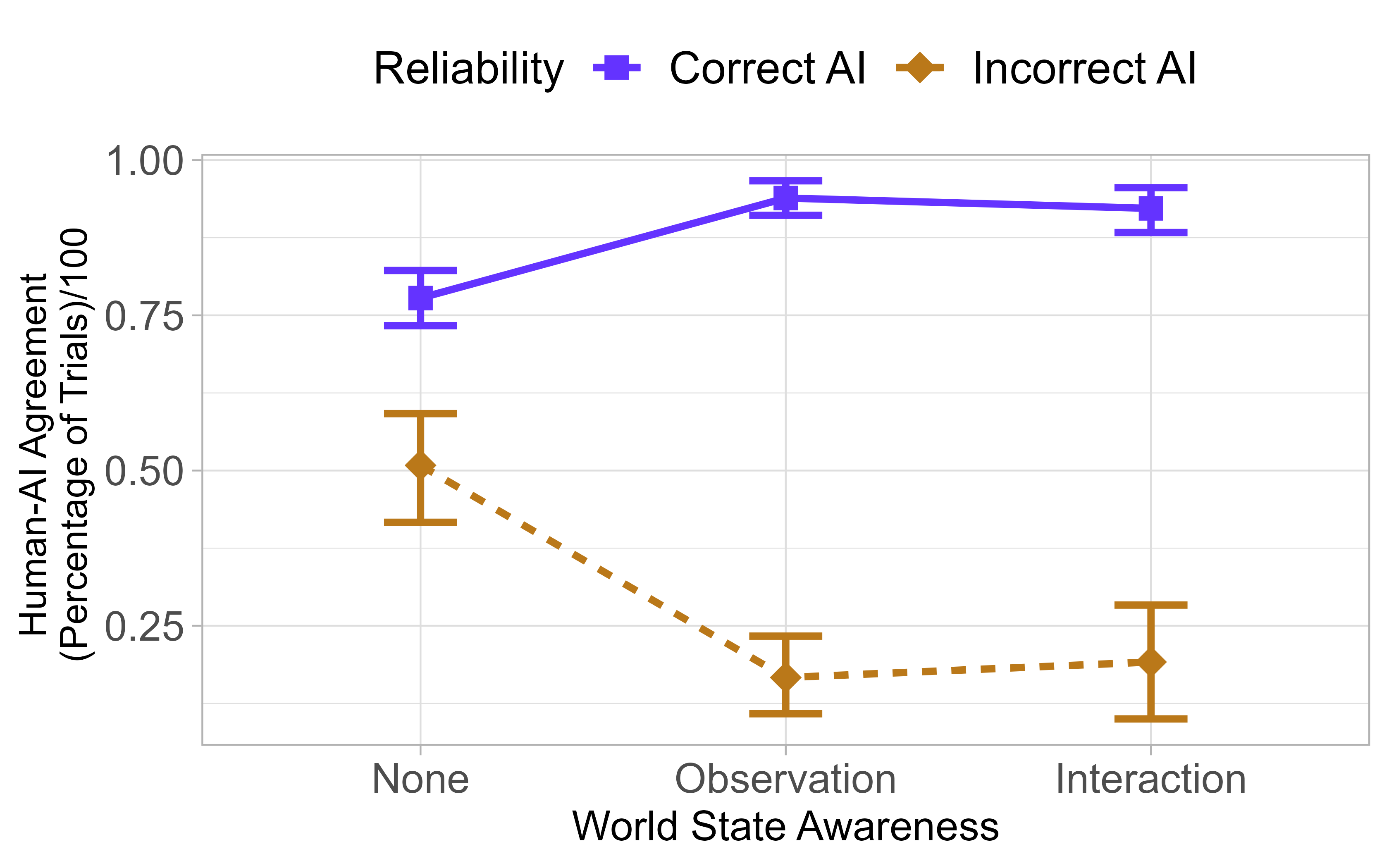}
    \caption{Final Decision Agreement [\%] Between Human and 60\% Accurate AI for both Trajectory State Awareness Levels}
    \label{fig:final_agr_exp2}
\end{figure}

\begin{table}[h]
\caption{\label{tab:final_agr_anova_exp2} ANOVA for Final Agreement (Phase 2)}%
\begin{tabular}{@{}llll@{}}
\toprule
 & df, error & F & P \\
\midrule
WSAwareness & 2, 82 & 2.597 & 0.0806    \\
TrajAwareness & 1, 82 & 0.937 & 0.3358    \\
Faith in Tech & 1, 82 & 2.888 & 0.0931   \\
Faith in Persons & 1, 82 & 0.270 & 0.6046   \\
WS-A:Traj-A & 2, 82 & 0.167 & 0.8462    \\
\botrule
\end{tabular}
\end{table}

As the final SSA could also be viewed as the initial decision agreement between the AI and the Human, we went further to understand how the trends for agreement were influenced by the reliability of the AI.
Figure \ref{fig:final_agr_exp2} depicts the team's final decision agreement (yellow diamond \#3 in Fig.~\ref{fig:task}).  In this figure, we collapsed across trajectory awareness for clarity. 
Ideally, when the AI is correct, the human would agree with it 100\% of the time; conversely when the AI is incorrect, that agreement would drop to 0\%. 
The average final agreement between the participant and the AI when the AI was correct (purple line in Fig.  \ref{fig:final_agr_exp2}) trended upward toward 100\% when the participant had access to world state information. 
We also see that the average final agreement between the participant and the AI when the AI was incorrect (brown line in Fig.  \ref{fig:final_agr_exp2}) trended downward toward 40\% as the participant's world state awareness increased.
This indicates that providing contextual information to the human improves their mental model of their AI partner (understanding that their AI teammate is limited in some capacity). 
In turn, this improvement in the human's mental model helps them determine when (and when not) to align with the AI's suggestion for the final decision.  
Further this indicates a bias toward AI agreement, particularly in the incorrect cases, that can be reduced, but not eliminated, with the introduction of world state.

%We performed a linear mixed effects analysis of the relationship between the final decision agreement between the participant and the AI and the  independent variables.
%We then conducted an ANOVA on this model (Table \ref{tab:final_agr_anova_exp2}).
None of the fixed effects in the linear-mixed effects model were statistically significant in predicting the final agreement between the team (Table \ref{tab:final_agr_anova_exp2}). However, when we performed a linear mixed effects analysis and ANOVA of the same metric specifically in scenarios in which the AI is incorrect, we found that World State Awareness was statistically significant in predicting final agreement between the team (Table \ref{tab:final_agr_incorrectAI}).

\begin{table}[h]
\caption{\label{tab:final_agr_incorrectAI} ANOVA for Final Agreement in Scenarios in which the AI is Incorrect (Phase 2)}%
\begin{tabular}{@{}llll@{}}
\toprule
 & df, error & F & P \\
\midrule
WSAwareness & 2,   82 & 8.115 & 0.0006   \\
TrajAwareness & 1,   82 & 2.471 & 0.1198    \\
Faith in Tech & 1,   82 & 2.193 & 0.1425   \\
Faith in Persons & 1,   82 & 0.300 & 0.5852  \\
WS-A:Traj-A & 2,   82 & 1.167 & 0.3165    \\
\botrule
\end{tabular}
\end{table}

\subsubsection{Subjective Metrics}

\begin{figure}[h]
    \centering
    \includegraphics[width=0.8\textwidth]{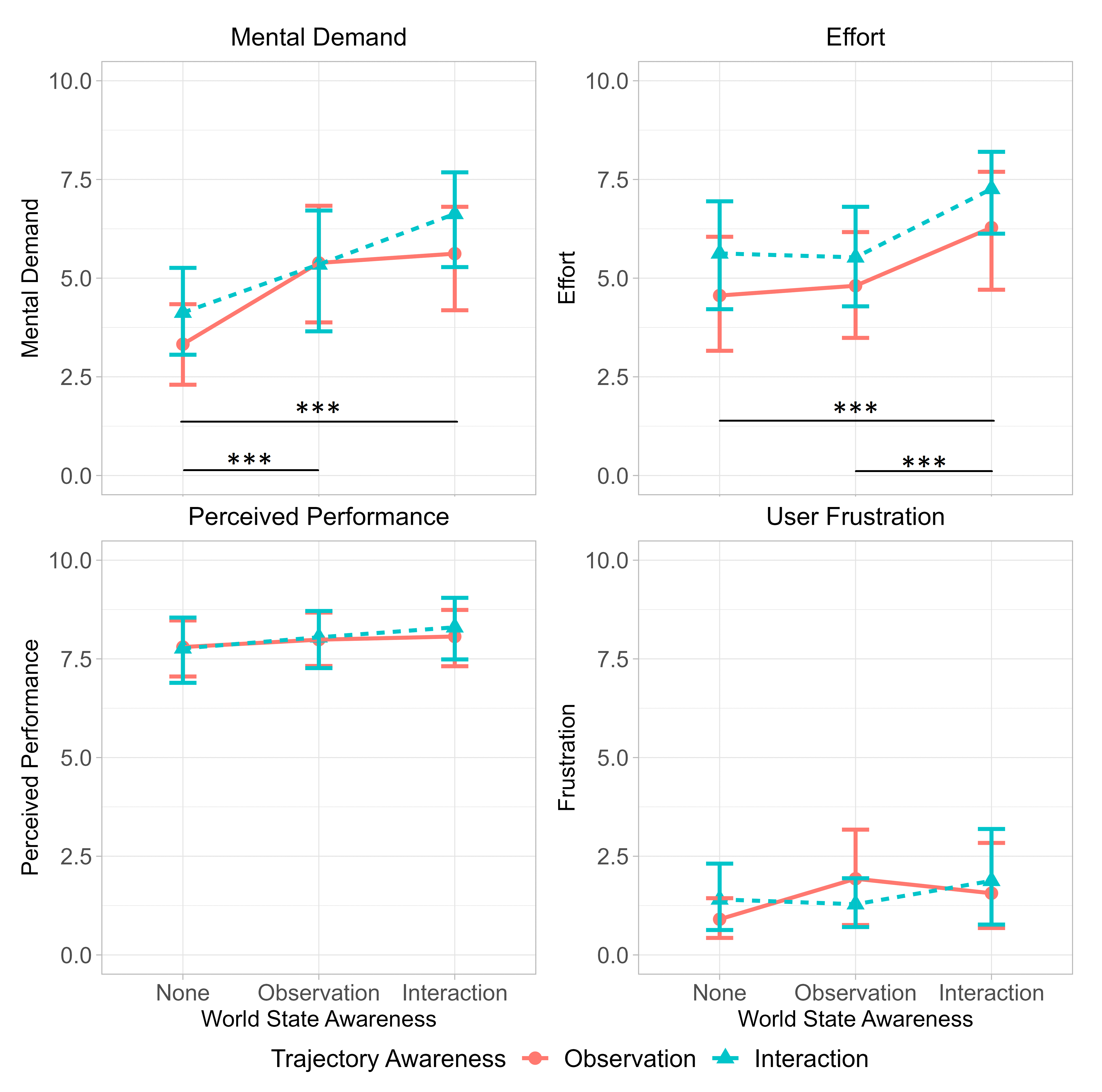}
    \caption{Composite Results of NASA TLX Questionnaire}
    \label{fig:exp2_tlx}
\end{figure}

As in the first experimental phase, we recorded the subjective mental workload of each participant using the NASA TLX assessment at the end of the experiment.  
Figure \ref{fig:exp2_tlx} depicts the results of the NASA TLX questionnaire for Experimental Phase 2.
%For all metrics except perceived performance, higher values indicate a higher workload on the participant.
%For perceived performance, higher values indicate a better perception of their own performance.
Similar to results from the first experimental phase, participants who had access to world state information reported more effort (F value = 3.5694) and mental demand (F value = 6.5426) required of them to complete the task. Both frustration (F value = 0.7146) and perceived performance (F value = 0.5273) were not significantly impacted by the world state  variable.
Results also show that those who interacted with the figures of merit (Trajectory Awareness Interaction) during the Trajectory Evaluation phase tended to experience more mental demand, effort, and frustration, but still perceived themselves as having high success in their performance.

\begin{figure}[h!]
    \centering
    \includegraphics[width=0.75\textwidth]{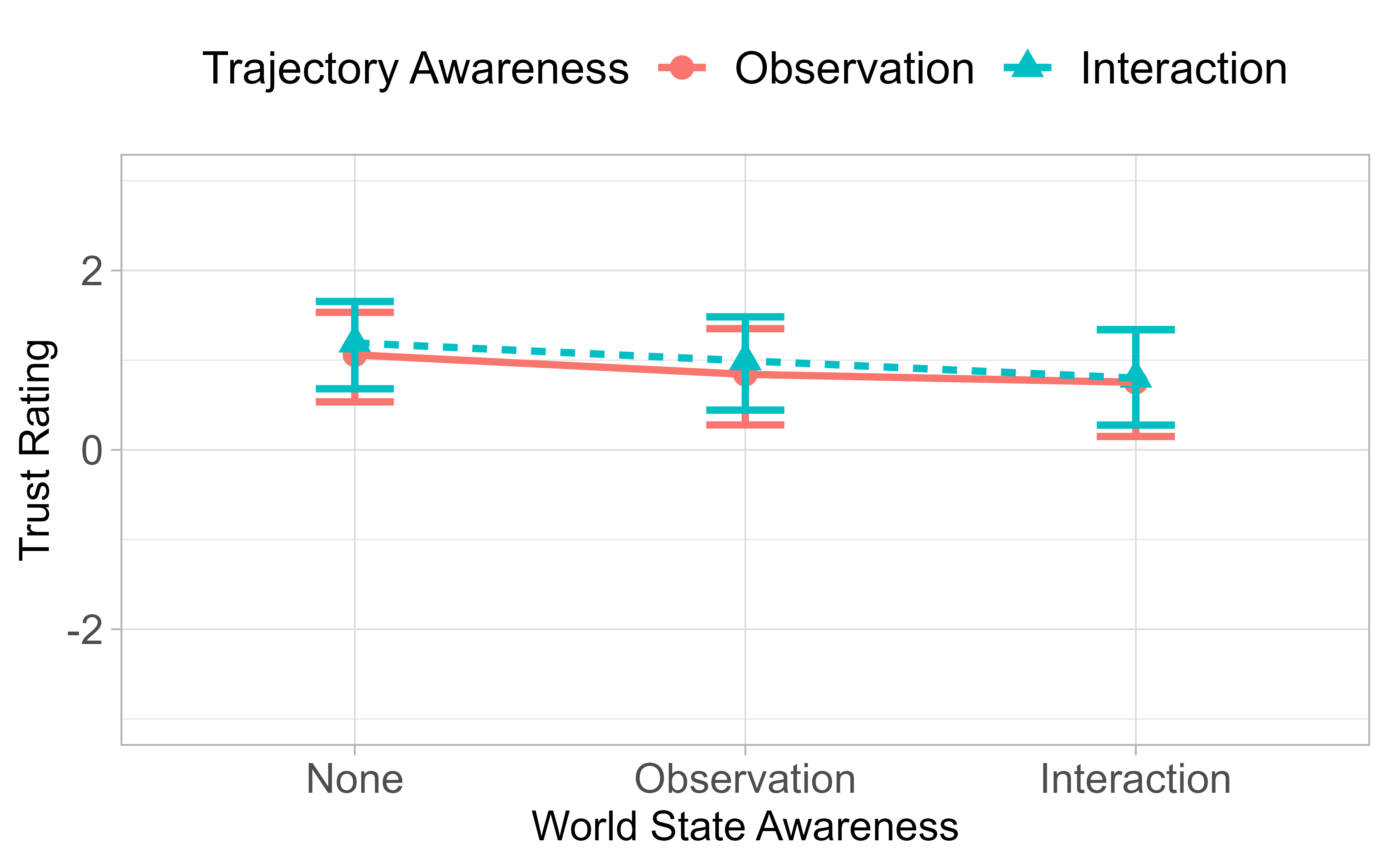}
    \caption{Post-Experiment i-THAu Metric of Human's Trust in AI's Capabilities}
    \label{fig:exp2_trust_capability}
\end{figure}

Similar to Phase 1, the measured dispositional trust from the pre-experiment i-THAu assessment was not statistically significant in predicting any objective measures (Tables \ref{tab:SSA_anova_exp2}-\ref{tab:final_agr_incorrectAI}), nor subjective measures such as cognitive workload or learned trust. Figure  \ref{fig:exp2_trust_capability} shows the human's trust in the AI's capabilities to complete the task measured in the post-experiment i-THAu assessment.  World State Awareness was not a statistically significant (p=0.4764) predictor of Capability Trust.  People were slightly more trusting of the AI’s capabilities when they interacted with the trajectory charts.

\section{Discussion}
The underlying premise of this study was to understand the importance of supporting human judgment as part of a more holistic cognitive process when designing recommender systems.  
We tested this by evaluating the impact of providing contextual information about the task environment, which we believed would increase the situation awareness of the human decision maker. 
Further, we aligned the contextual information specifically to match the information used by an AI recommender system in this task domain, so that we could remove any situation awareness discrepancies between the two agents. 
We compared this judgment support technique and shared situation awareness to the more commonly applied suggestion evaluation support technique of added interaction to determine its relative impact on the overall mission performance.  
We also added a final opportunity for the human to change their assessment to align with that of the AI recommender, to understand how providing judgment support and shared situation awareness might combat automation bias. 
Finally, we further assessed the impact of the additional tasks on mental workload and situational \& learned trust.  

The combined results from both phases indicate that providing contextual information of the task environment to humans improves their team performance with their AI partner, regardless of the accuracy of that partner.  
In the first experimental phase, the AI Mission Computer was always correct, so agreeing with the AI was equivalent to performing correctly.
The results from Phase 1 indicate that providing world state information increased shared situation awareness (Fig. \ref{fig:final_ssa}) and improved team performance. 
However, we see from Experimental Phase 2, (when the AI Mission Computer is 60\% accurate), that when the AI is incorrect in its suggestion, providing world state information helps to adjust the human's situation awareness, thereby aligning the team's SSA (Fig.~\ref{fig:final_ssa_exp2}). 
This indicates that providing contextual information is an effective technique in calibrating team's shared situation awareness, regardless of the autonomous agent's reliability.

World state awareness also appeared to protect against over reliance on automation/their AI teammate. 
The results of our baseline conditions show that those who do not have access to the world state conditions have a tendency to over-rely on their AI partner to make the correct decision (Figs. \ref{fig:final_agr_exp2} and \ref{fig:sway}).
This finding is crucial because, as seen from objective metrics (Figs. \ref{fig:final_agr} and \ref{fig:final_agr_exp2}) and subjective metrics (Figs. \ref{fig:exp1_tlx} and \ref{fig:exp2_tlx}), not seeing the world state information may result in acceptable final performance and less effort, mental demand, and frustration on the human's part, but ultimately leads to humans who are less confident in their own performance and a blind deference to an AI model that the human cannot understand, which as established, can lead to catastrophic results. 
Correspondingly, providing contextual information about the task environment aids the human in better understanding when their AI teammate is incorrect (Figs. \ref{fig:final_ssa_exp2}-\ref{fig:final_agr_exp2}, Table \ref{tab:final_agr_incorrectAI}), what to do about it, and improves their overall confidence in their own task performance (Fig. \ref{fig:exp2_tlx}).

Additionally, including the human in the Observation part of the decision making process by giving them contextual information increases their cognitive workload.
The TLX data from both experimental phases (Figs. \ref{fig:exp1_tlx} and \ref{fig:exp2_tlx}) indicate that more effort and mental demand is required from humans when given contextual information.
When working with a reliable AI partner, having contextual information ultimately makes the task less frustrating whereas frustration increases if the human is working with an unreliable AI partner. 
However, in both experimental phases, accessing the world state information raised the human's own confidence in their ability to perform the task.
Holistically, these workload results indicate that, although more is required of the human when given contextual information, the technique ultimately improves their objective performance as well as their own self-assessment in completing the task with their AI partner. 

The human's trust in the AI is not affected by the presence of contextual information, but is influenced by the reliability of the AI partner.
In Experimental Phase 1, there was little variation in the trust ratings of the AI's capabilities amongst the different levels of world state awareness (Fig. \ref{fig:exp1_trust_capability}).
However, in Experimental Phase 2, trust in the AI's capabilities decreased as world state awareness increased (Fig. \ref{fig:exp1_trust_capability}), although this downward trend was not statistically significant.
This finding, that humans are less trusting/forgiving of AI once they pick up on errors, is in line with current literature. 

Based on overall results, we can see that providing contextual information to the human and including them in the Observation stage of the decision making process can accurately calibrate team shared situation awareness, improve the human's overall task performance and their mental model of their AI teammate, and decrease over-reliance on automation without much added workload.

In an effort to support the evaluation of a suggested course of action, we introduced an interaction component during the Trajectory Evaluation phase.
The added interaction is meant to force the human to actively process each information source.
In Experimental Phase 1, we found that making sure humans understood the ``goodness" of the trajectory suggestion did not improve performance of the human significantly nor did it improve shared situation awareness between the teammates, which may contradict existing literature.
However, in Experimental Phase 2, those who interacted with the suggested trajectory's figures of merit had more accurate team SSA and final performance than those who merely observed the AI's suggestion.
Additionally, from both experimental phases, the TLX data trends overall indicate that the interaction component tended to lead to a worse user experience than merely observing the trajectory characteristics.
Since no real-world automation will be flawless and 100\% accurate in its outputs, this important finding indicates that adding interaction to AI outputs/suggestions is an effective technique for increasing SSA in real-world human-AI teams with unreliable AI team members, despite the poorer human experience.
However, it is important to note that this technique is not statistically significant in improving team SSA and final performance, so supporting the evaluation of a suggested course of action is not enough on its own to counteract the pitfalls of the black box phenomenon.

This study provides some evidence of the importance of situating a human decision maker, aligning the human's understanding of the world to match that of its teammates (in this case AI ones), and providing support for not just decision making, but also judgment elements of the cognitive process.  
The study however is limited by the simplistic nature of the task assessed. 
It also is limited in that the errors that the AI made were all risky and were designed to be detectable by the novice human decision makers.  
The participants here were also all novices and only worked in the team for a very short period of time - this may have limited the trust metric findings.  
Importantly, although we designed the experiments such that SSA should increase by providing contextual information, we cannot explicitly verify this because we do not measure the same metrics across treatments (e.g. we did not have a way to explicitly measure the situation awareness of those who did not see world state nor interact with figures of merit). 
But based on the final decisions made, we know that as world state awareness increases, the team's performance increases. 
Since we know from SMM literature that high performing teams utilize SMM, we can attribute this increase in performance to an increase in SSA, which we achieved through the technique of adding contextual information.
Finally, we cannot separate out the impact of SSA from our best guess as to the human's mental model. 

\section{Conclusions}
Results indicate that introducing transparency of the information used to generate trajectories accurately calibrates the shared situation awareness between the human and the AI (Figs. \ref{fig:final_ssa} and \ref{fig:final_ssa_exp2}).
While increasing transparency into the mission planning system by including the human in the Observation part of the process resulted in increased cognitive workload on the human's part, it also improved human robustness to overreliance (Figs. \ref{fig:sway} and \ref{fig:final_agr_exp2}), final team performance when the AI teammate was unreliable (Fig. \ref{fig:final_agr_exp2}), and the participants' confidence in their own performance (Fig. \ref{fig:exp1_tlx}).

Added interaction with the suggested solution led to poorer SSA (Fig. \ref{fig:final_ssa}), increased workload (Fig. \ref{fig:exp1_tlx}), and less trust in the AI teammate (Fig. \ref{fig:exp1_trust_capability}) when it was 100\% reliable.
While added interaction did seem to help calibrate SSA and trust in the AI when the AI was unreliable (Figs. \ref{fig:final_ssa_exp2} and \ref{fig:exp2_trust_capability}), it was not statistically impactful in these metrics, and therefore we cannot conclude that this technique is useful holistically.  

Our proposed technique of displaying contextual information about the task environment is broadly applicable to high-stakes decision support systems, in particular to improve human-AI shared situational awareness in systems built around \textit{black box} recommender systems. 
This work can be extended by testing if these findings hold when contextual information is relayed differently, in a different decision making domain, or how this technique affects human-AI team performance in conjunction with other transparency methods.

\bmhead{Acknowledgments}

This work was funded by Sandia National Laboratory (SNL) with Dr. Paul Schutte serving as Program Manager.  
This work is solely that of the authors and does not represent an official SNL position.

\bibliography{sn-bibliography}% common bib file

\end{document}